# Gyrokinetic full-f particle-in-cell simulations on open field lines with PICLS

M. Boesl, A. Bergmann, A. Bottino, D. Coster, E. Lanti, N. Ohana, and F. Jenko

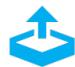

View Online    Export Citation    CrossMark

### ARTICLES YOU MAY BE INTERESTED IN

Approaching a burning plasma on the NIF
Physics of Plasmas **26**, 052704 (2019); https://doi.org/10.1063/1.5087256

Quantum hydrodynamics for plasmas—Quo vadis?
Physics of Plasmas **26**, 090601 (2019); https://doi.org/10.1063/1.5097885

Exact Solov'ev equilibrium with an arbitrary boundary
Physics of Plasmas **26**, 112503 (2019); https://doi.org/10.1063/1.5124138

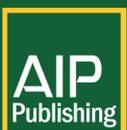

# Gyrokinetic full-f particle-in-cell simulations on open field lines with PICLS

M. Boesl,[1,a]] 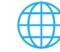 A. Bergmann,[1] A. Bottino,[1] D. Coster,[1] 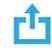 E. Lanti,[2] N. Ohana,[2] 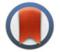 and F. Jenko[1]

AFFILIATIONS

[1]Max Planck Institut für Plasmaphysik, D-85748 Garching, Germany
[2]Ecole Polytechnique Fédérale de Lausanne, Swiss Plasma Center, CH-1015 Lausanne, Switzerland

[a)]Electronic mail: mathias.boesl@ipp.mpg.de

## ABSTRACT

While in recent years gyrokinetic simulations have become the workhorse for theoretical turbulence and transport studies in the plasma core, their application to the edge and scrape-off layer (SOL) region presents significant challenges. In particular, steep density and temperature gradients as well as large fluctuation amplitudes call for a "full-f" treatment. To specifically study problems in the SOL region, the gyrokinetic particle-in-cell (PIC) code PICLS has been developed. The code is based on an electrostatic full-f model with linearized field equations and uses kinetic electrons. Here, the well-studied parallel transport problem during an edge-localized mode in the SOL shall be investigated for one spatial dimension. The results are compared to previous gyrokinetic continuum and fully kinetic PIC simulations and show good agreement.

Published under license by AIP Publishing. https://doi.org/10.1063/1.5121262

## I. INTRODUCTION

In the plasma core, a region with closed field lines, gyrokinetic simulations have been carried out for several decades and proven to be able to correctly simulate turbulence.[1–8] Despite these promising results, the application of gyrokinetic codes to the plasma edge and scrape-off layer (SOL) region is hampered by various complications. Steep parallel gradients and wide ranges of spatial and temporal scales of plasma structures can be challenging for the assumptions taken in gyrokinetic theory. However, since edge and SOL effects are crucial for magnetic confinement, extending gyrokinetic simulations toward these regions is essential. In the present paper, we introduce a new gyrokinetic particle-in-cell (PIC) code called PICLS, which addresses the above challenges. Other gyrokinetic continuum[9,10] and PIC[11] codes already have been developed and started to contribute to the theoretical understanding of the SOL region, but still a great amount of research needs to be done. As a first test, PICLS is applied to the well-examined problem of parallel particle and energy transport caused by a transient type I edge-localized mode (ELM) in the SOL.

Type I ELMs (or "giant" ELMs) can be described as MHD instabilities that are triggered by steep pressure gradients in the plasma edge region (such as the H-mode pedestal) and lead to a loss of stored energy and a profile relaxation.[12] They can cause periodic energy outbursts into the SOL in the H-mode[13] and, due to the subsequent power loads on plasma facing materials, are a key issue for high-power tokamaks like ITER. These bursts can carry a significant portion of the stored energy.[14] In ITER, the eroded materials potentially require a higher replacement frequency, and eroded atoms within the plasma chamber deteriorate the energy gain. Studies showed that hundreds of ELMs are expected to occur in only one single ITER discharge.[15] With such a high number of plasma wall interactions via ELMs, their suppression is of the highest importance to limit the damage on plasma facing components (such as a divertor or main chamber wall). Apart from experimental studies for ELM suppression, an accurate prediction of heat transport in the SOL for future devices via simulation is also required.

The ELM heat pulse problem was already studied within several simulation efforts, especially in the 1D1V (one spatial and one velocity domain) case. Initially, parallel propagation of ELM heat pulses with different temperatures, densities, energies, and durations was simulated via a fully kinetic collisional PIC model.[14] The ELM pulses in the midplane of the SOL were derived from JET experiments and modeled as hot plasma sources in the center of a 1D simulation domain. By benchmarking their simulation data against the experiment, the quantity range of the fraction of energy that was deposited before the heat flux peak in the JET measurement could be correctly predicted. Although the ELM model is quite reduced, it is viable to simulate divertor heat fluxes and parallel transport in the SOL and is in good agreement with experiments. In a more recent study, the single central source model was studied with fully kinetic PIC, continuum (Vlasov), and fluid simulations and successfully benchmarked against the experiment.[16] Also, between the different codes, a consistent outcome could be achieved.[14,16,17]

In very recent works, the same problem was simulated with gyrokinetic continuum codes and the results of the previous works could be reproduced.[18–20] Additionally, both implemented logical-sheath boundary conditions,[21] which are designed to provide the effects of a Debye sheath, while not actually having to resolve it. Hence, the combination of gyrokinetics and a logical sheath delivers a significant speed up, compared to kinetic codes, since the restriction toward time steps of $\approx \omega_{ce}^{-1}$ (electron cyclotron frequency) and spatial resolutions of $\approx \lambda_{De}$ (Debye length) is lifted. Our implemented model can use time steps and grid cells that are several orders of magnitude larger than required in kinetic models that need to resolve the Debye length and the plasma frequency. Compared to fluid models, this approach is surely computationally heavier, but also ensures that kinetic effects are taken into account.

Unlike the previous gyrokinetic simulations, we apply a PIC model to simulate the same ELM heat pulse problem. This model is implemented in the newly developed PICLS code, which is designed to perform gyrokinetic SOL simulations. Its numerical core is heavily based on numerical techniques implemented in ORB5[22] and GK-Engine.[23] Since a typical $\delta f$ model for the distribution function (splitting in a constant background part and a perturbation) is not useful in the SOL anymore, a full-f model with a linearization in the field part is implemented in PICLS. For the simulations in this paper, we focus on long wavelengths in the drift-kinetic limit. The electrostatic potential is calculated via the polarization equation, with the help of B-spline finite-elements for the charge deposition and the field solver. For the time integrator, a fourth-order Runge-Kutta algorithm is used and it is parallelized with a hybrid OpenMP/MPI setup.

With this simplified problem, the implemented sheath physics is tested and its consistency with previous simulations can be benchmarked. The one dimensional setup is also an important step toward studying basic plasma physics phenomena of real linear devices, such as large plasma device (LAPD) at UCLA.[24] Studies on this machine were previously performed by the full-f gyrokinetic continuum codes Gkeyll[9] and GENE.[10]

In Sec. II, we describe the general electrostatic gyrokinetic equations implemented in PICLS and its specific formulations for the 1D heat pulse problem. The numerical implementation details for our PIC approach and the logical sheath boundaries are introduced in Sec. III. The heat pulse simulation setup is described in Sec. IV together with the corresponding results in Sec. V. The conclusions and an outlook on future efforts are shown in Sec. VI.

## II. PHYSICAL MODEL

Throughout this paper, we use a low-frequency, electrostatic gyrokinetic model with kinetic electrons. Finite-Larmor radius effects are neglected due to the nature of the single spatial dimension ELM pulse problem we investigate. However, Larmor-radius effects and gyroaveraging are already implemented within the code for a future higher dimensional extension.

### A. Basic equations

Here, we want to derive the required set of equations in a general three dimensional case and subsequently reduce these to the specific set required for the applied simulations in one spatial dimension. The starting point is a total gyrokinetic Lagrangian in centimeter-gram-second (CGS) units,[25]

$$L = \sum_p \int dW_0 dV_0 f_p(\mathbf{Z}_0, t_0) L_p(\mathbf{Z}(\mathbf{Z}_0, t_0; t), \dot{\mathbf{Z}}(\mathbf{Z}_0, t_0; t), t) + \int dV \frac{E^2 - B_\perp^2}{8\pi}, \quad (1)$$

where $\mathbf{Z} \equiv (\mathbf{R}, v_\parallel, \mu, \theta)$, and $dW$ and $dV$ stand for the volume elements in velocity and physical space, respectively. With the gyrocenter position $\mathbf{R}$ and the velocity variables $v_\parallel$ (velocity parallel to the magnetic field), $\mu = m_p v_\perp^2/(2B)$ (magnetic moment) and $\theta$ (gyroangle). $f(\mathbf{Z}_0) = f_p$ is the distribution function for the species $p$ at initial time $t_0$.

With the following definitions:

$$dW = \frac{2\pi}{m_p} B_\parallel^* dv_\parallel d\mu, \quad (2)$$

$$dV = J(X, Y, Z) dX dY dZ, \quad (3)$$

$$d\Lambda = dV dW, \quad (4)$$

$$B_\parallel^* \equiv \mathbf{B}^* \cdot \mathbf{b} = B + \frac{c m_p}{e_p} v_\parallel \nabla \times \mathbf{b} \cdot \mathbf{b}, \quad (5)$$

$$\mathbf{B}^* = \mathbf{B} + \frac{m_p c}{e_p} v_\parallel \nabla \times \mathbf{b}, \quad (6)$$

where $m_p$ is the mass and $e_p$ the charge of species $p$, we can further condense the total Lagrangian to

$$L = \sum_p \int d\Lambda L_p f_p + \int dV \frac{E^2 - B_\perp^2}{8\pi}. \quad (7)$$

Here, the Lie-transformed low-frequency particle Lagrangian $L_p$ is used,[26,27]

$$L_p \equiv \left(\frac{e}{c}\mathbf{A} + m_p v_\parallel \mathbf{b}\right) \cdot \dot{\mathbf{R}} + \frac{m_p c}{e_p} \mu \dot\theta - H_p, \quad (8)$$

with $\mathbf{A}$ being the background vector potential. Components with a perpendicular subscript lie within the plane perpendicular to the magnetic background field $\mathbf{B}$. The (electrostatic) Hamiltonian used for the subsequent derivations in our case is the following (but also other choices would be possible):

$$H_p = m_p \frac{v_\parallel^2}{2} + \mu B + e_p J_{p,0} \phi - \frac{m_p c^2}{2B^2} |\nabla_\perp \phi|^2, \quad (9)$$

which can also be written as

$$H_p \equiv H_{p,0} + H_{p,1} + H_{p,2}, \quad (10)$$

$$H_{p,0} = \frac{m_p v_\parallel^2}{2} + \mu B, \quad (11)$$

$$H_{p,1} = e_p J_{p,0} \phi, \quad (12)$$

$$H_{p,2} = -\frac{m_p c^2}{2B^2} |\nabla_\perp \phi|^2. \quad (13)$$

In $H_{p,1}$, the gyroaveraging operator $J_{p,0}$ is applied to the potential $\phi$. The operator $J_{p,0}$ applied to an arbitrary function $\psi$ in configuration space is defined by

$$(J_0\psi)(\mathbf{R},\mu) = \frac{1}{2\pi}\int_0^{2\pi}\psi(\mathbf{R}+\boldsymbol{\rho}(\theta))\,d\theta, \tag{14}$$

with $\boldsymbol{\rho}$ being the vector from the guiding center position to the particle position.

Now, by substituting Eq. (8) into Eq. (7), the required total gyrokinetic particle Lagrangian can be achieved as follows:

$$L = \sum_p \int d\Lambda \left(\left(\frac{e_p}{c}\mathbf{A} + m_p v_\parallel \mathbf{b}\right)\cdot\dot{\mathbf{R}} + \frac{m_p c}{e_p}\mu\dot\theta - H_p\right)f_p$$
$$+ \int dV\frac{E^2 - B_\perp^2}{8\pi}. \tag{15}$$

According to gyrokinetic (GK) theory, the resulting Lagrangian (15) can be further approximated without losing energetic consistency and self-consistency of the final equations.[25] In this work, the widely used quasineutrality approximation and linearized polarization approximation are applied.

For the quasineutrality approximation, the $E^2$ term in the free-field term is ordered small compared to the so-called $E\times B$ term, which corresponds to the second order $\phi$ term in the Hamiltonian. To see this, the Debye length squared $\lambda_{\rm De}^2 \equiv \frac{k_B T_e}{4\pi n_p e^2}$ and the squared ion sound Larmor radius $\rho_s^2 \equiv \frac{k_B T_e m_i c^2}{e^2 B^2}$ are introduced. In fusion plasmas, normally the ion Larmor radius, in general, is much larger than the Debye length, and one obtains

$$\frac{\rho_s^2}{\lambda_{\rm De}^2} = \frac{4\pi n_p m_p c^2}{B^2} = \frac{c^2}{v_a^2} \gg 1, \tag{16}$$

with $v_a$ being the Alfvén velocity. This relationship is now used in the sum of the $E^2$ term in the free fields and the second order $\phi$ term in the Hamiltonian,

$$\int dV\frac{E^2}{8\pi} + \int d\Omega f\frac{m}{2}\frac{c^2}{B^2}|\nabla_\perp\phi|^2 = \frac{1}{8\pi}\int dV\left[E_\parallel^2 + \left(1+\frac{\rho_s^2}{\lambda_{\rm De}^2}\right)|\nabla_\perp\phi|^2\right]. \tag{17}$$

The $E_\parallel^2$ term is even smaller than the perpendicular part, and hence, the whole $E^2$ term can be neglected.

In a second step, the Lagrangian is further approximated, by assuming, that only the $(H_{p,0}+H_{p,1})$ part of the Hamiltonian multiplies with the full distribution function $f$ and $H_{p,2}$ is linearized by multiplying with a time-independent equilibrium distribution function $f_{M,p}$,

$$L = \sum_p \int d\Lambda \left(\left(\frac{e_p}{c}\mathbf{A} + m_p v_\parallel \mathbf{b}\right)\cdot\dot{\mathbf{R}} + \frac{m_p c}{e_p}\mu\dot\theta - H_{p,0} - H_{p,1}\right)f_p$$
$$+ \sum_p \int d\Lambda \frac{m_p c^2}{2B^2}|\nabla_\perp\phi|^2 f_{M,p} - \int dV\frac{B_\perp^2}{8\pi}. \tag{18}$$

The linearized field part is known as the linearized polarization approximation. Thus, the $H_{p,2}$ term only acts on the field equations and does not contribute to the drift motion. For a detailed overview of both applied Lagrangian approximations, we refer to Bottino and Sonnendrücker.[26]

Since all our simulations are performed in the electrostatic limit, electromagnetic perturbations are also neglected, by setting $A_\parallel = 0$, which implies $B_\perp^2 = 0$, and therewith the $\int dV\frac{B_\perp^2}{8\pi}$ term in the field part is neglected. Our final electrostatic total Lagrangian is thus

$$L = \sum_p \int d\Lambda \left(\left(\frac{e_p}{c}\mathbf{A} + m_p v_\parallel \mathbf{b}\right)\cdot\dot{\mathbf{R}} + \frac{m_p c}{e_p}\mu\dot\theta - H_{p,0} - H_{p,1}\right)f_p$$
$$+ \sum_p \int d\Lambda \frac{m_p c^2}{2B^2}|\nabla_\perp\phi|^2 f_{M,p}. \tag{19}$$

The particle equations of motion can be obtained by taking the functional derivative of $L_p$ with respect to $\mathbf{Z}$, and from this, the Euler-Lagrange equations for $L_p$ can be derived.[25,26] These describe the drift motion of the gyrocenters and can be written as

$$\frac{d}{dt}\frac{\delta L_p}{\delta \dot{\mathbf{Z}}} = \frac{\delta L_p}{\delta \mathbf{Z}}. \tag{20}$$

Following Bottino and Sonnendrücker,[26] one can now calculate all required derivatives of $L_p$. For the $\theta$ derivatives $\frac{\partial L_p}{\partial\theta}=\mu, \frac{\partial L_p}{\partial\dot\theta}=0$ can be calculated. Thus, the Euler-Lagrange equation for $\theta$ delivers $\frac{d\mu}{dt}=0$ and shows that $\mu$ is an exact invariant. An evolution equation for $\theta$ is not needed, since the dependence on $\theta$ has been removed from the Lagrangian. For $\mathbf{R}$ and $v_\parallel$, the Euler-Lagrange equations yield

$$\dot{\mathbf{R}} = v_\parallel\frac{\mathbf{B}^*}{B_\parallel^*} + \frac{c}{e_p B B_\parallel^*}\mathbf{B}\times[\mu\nabla B + e_p\nabla J_{p,0}\phi],$$
$$\dot v_\parallel = -\frac{\mathbf{B}^*}{B_\parallel^*}\frac{1}{m_p}\cdot[\mu\nabla B + e_p\nabla J_{p,0}\phi]. \tag{21}$$

To derive the polarization equation, or GK Poisson equation, the functional derivative of $L$ needs to be calculated with respect to the electrostatic potential $\phi$ and set to 0. The only dependence on $\phi$ in the first term of Eq. (19) is in the $H_{p,1}$ part. The gyroaverage operator $J_{p,0}$ is a linear operator of $\phi$, and thus, we get

$$\frac{\delta L}{\delta\phi}\delta\phi = -\sum_p\left(\int d\Lambda e_p J_{p,0}(\delta\phi)f \right.$$
$$\left. + \int dV\frac{m_p c^2}{B^2}\nabla_\perp\phi\cdot\nabla_\perp\delta\phi f_{M,p}\right) = 0. \tag{22}$$

This form is also called the weak form of the polarization equation, which is solved in the code. The strong form results from using a Hermitian $J_{p,0}$,

$$\int\phi J_{p,0}(f)dW = \int f J_{p,0}(\phi)dW, \tag{23}$$

and from applying Green's formula on the second integral of (22). It is also assumed that $\phi$ vanishes at the boundary and $B_\parallel^*$ is taken out of $dW = \frac{2\pi}{m_p}B_\parallel^* dv_\parallel d\mu$. With this, Eq. (22) becomes

$$-\sum_p\int dV\delta\phi\int dW\left(e_p J_{p,0}f + \frac{1}{B_\parallel^*}\nabla_\perp\left(\frac{m_p c^2}{B^2}B_\parallel^* f_{M,p}\nabla_\perp\phi\right)\right) = 0. \tag{24}$$

Now, by noting that in the second term the spatial gradient and the integral with respect to $dW$ commute, we can perform the velocity integral over the Maxwellian distribution. Since the choice of $\delta\phi$ is arbitrary, we get

$$-\sum_p \nabla_\perp \frac{n_{p,0} m_p c^2}{B^2} \nabla_\perp \phi = \sum_p \int dW e_p J_{p,0} f. \quad (25)$$

With the definition $n_{p,0}$ for the density of the equilibrium Maxwellian $f_M$,

$$n_{p,0} = \int dW f_{M,p}. \quad (26)$$

Note that $J_{p,0} = 1$ for electrons, which means that electron finite Larmor radius effects are neglected, due to their small Larmor radius. The resulting equation is linear and has the form $\sum_p e n_p = 0$ with the particle charge density $e n_p$, which is a quasineutrality condition.

As shown in Shi,[19] a simplified polarization equation can be obtained. This is derived from a dispersion relation resulting from linearizing Eq. (25) and the Vlasov equation $\frac{\partial F_p}{\partial t} = \{H_p, F_p\}$, Fourier transforming both in time and space and assuming that $q_e = q_i$. By introducing the shielding factor $s_\perp(z,t) = k_\perp^2(z)\epsilon_\perp(z,t)$, with $\epsilon_\perp = \sum_p \frac{n_{p,0} m_p c^2}{B^2}$, the modified polarization equation can be written as

$$s_\perp(z,t)(\phi - \langle\phi\rangle) = \sum_p \int dW e_p J_{p,0} f, \quad (27)$$

where the flux-surface-averaged, dielectric-weighted potential

$$\langle\phi\rangle = \frac{\int dz s_\perp \phi}{\int dz s_\perp} \quad (28)$$

is used. In Eq. (27), the flux-surface-averaged potential $\langle\phi\rangle$ is subtracted off $\phi$ to maintain gauge invariance of the modified polarization equation. This choice is also taken due to the applied logical sheath boundary conditions. Specifically, in the chosen 1D case, the net guiding center charge vanishes, $\int \sigma_{tot} dz = \int \sum_p \int dW e_p J_{p,0} f dz = 0$, since the net flux is set to $j_\parallel = 0$. Applying the integral over $dz$ on the left-hand-side of Eq. (27) also shows that the polarization charge density averages to 0. However, for the calculation of the E-field, this subtraction of $\langle\phi\rangle$ can be neglected and has no influence on the gyrokinetic equations of motion. Hence, we achieve the modified polarization equation used in our model as

$$s_\perp(z)\phi(z) = k_\perp^2(z) \sum_p \frac{n_{p,0} m_p c^2}{B^2} \phi(z) = \sum_p \int dW e_p J_{p,0} f. \quad (29)$$

In the PIC algorithm with finite-element discretization, this means that the Poisson matrix has no longer to be solved, but the mass matrix with an additional factor of $k_\perp^2$.

For the calculation of the conserved energy of Eq. (27), we refer to Shi et al.[18] In this derivation, the $\langle\phi\rangle$ term would actually have to be taken into account. However, in this work, we want to show the conserved energy for the original unmodified polarization equation (25), which is relevant for future more general studies. Here, by deriving the equations from the not directly time-dependent Lagrangian density, the energy is automatically a conserved quantity. Adjustments on the Lagrangian and Hamiltonian level thus conserve the energy, whereas dropping terms at a later stage can result in a violation of energy conservation.[26,28] In our electrostatic case, the total conserved energy is[29]

$$\mathcal{E}_{tot} = \sum_p \int dW dV H_p f_p = \mathcal{E}_k + \mathcal{E}_f, \quad (30)$$

where the particle energy (or total kinetic energy) $\mathcal{E}_k$ has the definition

$$\mathcal{E}_k = \sum_p \int dW dV H_{p,0} f_p. \quad (31)$$

The definition for the field energy is

$$\mathcal{E}_f = \sum_p \int dW dV H_{p,1} f_p + \sum_p \int dW dV H_{p,2} f_{M,p}, \quad (32)$$

$$= \sum_p \int dW dV e_p J_{p,0} \phi f_p - \sum_p \int dW dV \frac{m_p c^2}{2B^2} |\nabla_\perp \phi|^2 f_{M,p}. \quad (33)$$

Considering the polarization equation (25), multiplying by $\phi$ and integrating over space give

$$-\sum_p \int dV dW \left(\nabla_\perp \frac{f_{p,0} m_p c^2}{B^2} \nabla_\perp \phi\right) \phi = \sum_p \int dW dV e_p J_{p,0} f \phi. \quad (34)$$

Now, we integrate by parts, and use the Hermiticity of $J_{p,0}$ and divide by 2, to obtain

$$\sum_p \int dV dW \frac{f_{p,0} m_p c^2}{2B^2} |\nabla_\perp \phi|^2 = \sum_p \int dW dV e_p J_{p,0} \phi f_p. \quad (35)$$

Inserting this result in the second term of Eq. (33) gives

$$\mathcal{E}_f = \sum_p \frac{1}{2} \int dW dV e_p J_{p,0} \phi f_p. \quad (36)$$

Thus, the total conserved energy of the system can be written as

$$\mathcal{E}_{tot} = \mathcal{E}_k + \mathcal{E}_f = \sum_p \int dW dV H_{p,0} f_p + \sum_p \frac{1}{2} \int dW dV e_p J_{p,0} \phi f_p, \quad (37)$$

with $H_{p,0} = \frac{1}{2} m_p v_\parallel^2 + \mu B$ in the kinetic part.

### B. Equations in 1D slab geometry

The equations of motion in slab geometry can be derived from Eq. (21) for the 3D case straightforward,

$$\dot{\mathbf{R}} = v_\parallel \mathbf{b} + \frac{c}{e_p} \mu \frac{\mathbf{B} \times \nabla B}{B^2} + c \frac{\mathbf{B} \times \nabla J_{p,0} \phi}{B^2},$$
$$\dot{v}_\parallel = -\frac{\mu}{m_p} \mathbf{b} \cdot \nabla B - \frac{e_p}{m_p} \mathbf{b} \cdot \nabla J_{p,0} \phi, \quad (38)$$

by using that $\mathbf{B}^* = \mathbf{B}$ and $B_\parallel^* \equiv \mathbf{B}^* \cdot \mathbf{b} = B$ in slab geometry. The Hamiltonian of the system remains the full-f gyrokinetic Hamiltonian of Eq. (9).

Within this work, the 1D1V versions of these equations are required to evolve the markers according to the PIC algorithm applied. The B-field in the 1D case is parallel to the z-direction of the domain, and in the case of the studied 1D1V ELM heat-pulse problem, $B = \text{const}$. Thus, in the $\dot{\mathbf{R}}$ part of Eq. (38), the $\mathbf{B} \times \nabla B$ and the $\mathbf{B} \times \nabla J_{p,0}$ terms cancel out. For the 1 V case, $\mu = 0$ also applies, and thus, the first term in the $\dot{v}_\parallel$ equation also disappears and hence further simplifies the equations. In the 1D case, Eq. (38) can thus be rewritten as

$$\dot{\mathbf{R}} = v_\parallel \mathbf{b},$$
$$\dot{v}_\parallel = -\frac{e_p}{m_p} \mathbf{b} \cdot \nabla J_{p,0} \phi. \tag{39}$$

The modified polarization equation (29) is unchanged in slab geometry, with $J_{p,0} = 1$ in the 1D case.

### C. Normalization scheme

We choose the centimeter-gram-second (or CGS) system of units because in this case, for electromagnetic studies, the $E$- and the $B$-field have the same units. To address these units within the simulations, the equations are normalized before numerical solutions are calculated. The full set of normalization equations can be written as

$$t = \tau \bar{t}, \quad R = L\bar{x}, \quad v_\parallel = V\bar{v}_\parallel, \quad k_\perp = \bar{k}_\perp / \rho_{ce},$$
$$e_p = e\bar{Z}_p, \quad m_p = M\bar{m}_p, \quad T_p = k_B T \bar{T}_p,$$
$$E \equiv \nabla J_{p,0}\phi = E_{\text{norm}} \bar{E} = \frac{k_B T}{eL} \bar{E},$$
$$B = B_{\text{norm}} c\bar{B} = \frac{c}{V} \frac{k_B T}{eL} \bar{B}, \quad \mu = \mu_{\text{norm}} \bar{\mu} = \frac{eLV^3 M}{ck_B T} \bar{\mu}.$$

We normalize $B/c$ rather than $B$; practically, this is similar to rescaling the field by $c/V$. Hence, $c$ disappears from the equations. In the code, $B$ and $T$ are initialized by their input values $B_0$ and $T_0$. Additionally, the spatial scale $L$ is set to the electron gyroradius,

$$\rho_{ce} = \sqrt{k_B T/m_e}/\frac{c}{eB_0}. \tag{40}$$

Hence, $\tau$ becomes the inverse electron cyclotron frequency $w_{ce}^{-1} = \frac{m_e c}{eB_0}$. The other reference quantities applied comprise the electron charge $e$, the electron mass $M$, and the initial thermal electron velocity $V = \sqrt{k_B T/m_e}$.

$k_\perp$ is normalized to the inverse of the electron cyclotron radius $1/\rho_{ce}$. Since $l_{\text{norm}} = \rho_{ce}$ is set in our simulations, the normalized form of the Fourier transformed polarization equation (29) can be obtained as

$$\sum_p \bar{k}_\perp^2 \frac{n_{p,0} M \bar{m}_p}{\bar{B}^2 B_{\text{norm}}^2} \phi(z) = \sum_p \int dW e \bar{Z}_p J_{p,0} f. \tag{41}$$

Entering the normalization factors into the equations of motion (39) for 1D yields

$$\frac{\partial \bar{z}}{\partial \bar{t}} = \bar{v}_\parallel,$$
$$\frac{\partial \bar{v}_\parallel}{\partial \bar{t}} = \left(-\frac{e\bar{Z}_p}{M\bar{m}_p} \bar{E}_z E_{\text{norm}}\right) \frac{\tau}{V}. \tag{42}$$

Our choice of the normalization scheme is primarily taken to easily perform a-dimensional scans in the future (e.g., on $\rho^*$ or $\nu^*$). However, other normalization schemes (or no normalization, such as SI or CGS systems of unit) are also possible. The choice is not decisive for numerical calculations, since variations of magnitudes can be easily handled by double precision numbers.

## III. NUMERICAL SETUP

As mentioned in Sec. I, for the simulations within this work, we use the newly developed PICLS code, which is a PIC code designed to perform gyrokinetic open-field-line simulations. The numerical function discretization is performed with a finite-element scheme, and the sheath at the domain boundary is implemented via a logical sheath model.

### A. Discretization of equations

In PIC codes, the particle distribution function $f(x)$ is represented by discrete markers. For the full-$f$ representation for the 3D space, 1D velocity space—where the whole distribution is simulated—the particle distribution function thus becomes

$$f(\mathbf{R}, v_\parallel, t) = \sum_{n=1}^{N} w_n(t) \delta(\mathbf{R} - \mathbf{R}_n(t)) \delta(v_\parallel - v_{\parallel n}(t)), \tag{43}$$

with $N$ being the number of markers, $w_n$ the marker weights, $\mathbf{R}_n$ their position, and $v_{\parallel n}$ their parallel velocity. The magnetic moment $\mu$ and the gyroangle $\theta$ can be neglected for this consideration. By defining $N_{\text{ph}} = \int n_0(\mathbf{R}) d\mathbf{R}$ as the initial number of physical particles, the weights are uniformly initialized with

$$w_n = \frac{N_{\text{ph}}}{N} \tag{44}$$

for all markers. The weights for the full-$f$ case are constant and do not evolve in time,

$$\frac{d}{dt} w_n = 0. \tag{45}$$

A key task of PIC codes is to solve the potential for a given particle distribution at each time step and from this calculate the electrical field. The $E$-field is required to advance the particles in the following time step. For the polarization equation (25), this means that the discretized form of $f(x)$ from Eq. (43) needs to be inserted into the polarization equation (25),

$$-\sum_p \nabla_\perp \frac{n_{p,0} m_p c^2}{B^2} \nabla_\perp \phi = \sum_p e_p \sum_{n=1}^{N} w_{p,n}(t) \delta(\mathbf{R}_p - \mathbf{R}_{p,n}), \tag{46}$$

with $p$ being the index for summing over particle species and $N$ the number of markers. Now, both sides are multiplied with a test function $\psi$ and integrated over the whole spatial domain $V$ to get

$$\int_V \psi \left(-\sum_p \nabla_\perp \frac{n_{p,0} m_p c^2}{B^2} \nabla_\perp \phi\right) d\mathbf{R} = \sum_p e_p \sum_{n=1}^{N} w_{p,n}(t) \psi(\mathbf{R}_n). \tag{47}$$

Integrating by parts and applying the divergence theorem simplify the left-hand side to

$$\int_V \psi \left(\sum_p \nabla_\perp \left(\frac{n_{p,0} m_p c^2}{B^2} \nabla_\perp \phi\right)\right) d\mathbf{R}$$
$$= \underbrace{\int_{\partial V} \left(\sum_p \psi \frac{n_{p,0} m_p c^2}{B^2} \nabla_\perp \phi\right) d\sigma}_{=0}$$
$$- \int_V \left(\sum_p \frac{n_{p,0} m_p c^2}{B^2} \nabla_\perp \psi \nabla_\perp \phi\right) d\mathbf{R}. \tag{48}$$

Here, the fact that the integral over the domain boundary is zero is used. Inserting this in Eq. (47) then yields

$$\sum_p \int_V \frac{n_{p,0} m_p c^2}{B^2} \nabla_\perp \psi \nabla_\perp \phi \, d\mathbf{R} = \sum_p e_p \sum_{n=1}^N w_{p,n}(t) \psi(\mathbf{R}_n). \quad (49)$$

This equation can now be discretized with the help of the finite-element method. Therefore, to represent the potential $\phi$, we use B-spline basis functions and obtain

$$\phi(\mathbf{R}) = \sum_\mu \hat{\phi}_\mu \Lambda_\mu^k(\mathbf{R}), \quad (50)$$

where $\mu$ in the 1D case is a single index and goes from $\mu = 1, \ldots, N_z$ the number of grid cells and $\Lambda_\mu^k(\mathbf{R})$ is the $\mu$th B-spline of degree $k$. For further and more detailed information on B-splines, we refer to De Boor.[30] Also, by choosing $\psi(\mathbf{R}) = \Lambda_\beta^k(\mathbf{R})$ for the test function, we can rewrite Eq. (49) as follows:

$$\sum_\mu \hat{\phi}_\mu \sum_p \int_V \frac{n_{p,0} m_p c^2}{B^2} \nabla_\perp \Lambda_\beta^k(\mathbf{R}) \nabla_\perp \Lambda_\mu^k(\mathbf{R}) d\mathbf{R}$$
$$= \sum_p e_p \sum_{n=1}^N w_{p,n}(t) \Lambda_\beta^k(\mathbf{R}_n). \quad (51)$$

The matrix $P_{\mu\beta}$ and the right-hand-side $\rho_\beta$ can now be defined in the following way:

$$P_{\mu\beta} = \sum_p \int_V \frac{n_{p,0} m_p c^2}{B^2} \nabla_\perp \Lambda_\beta^k(\mathbf{R}) \nabla_\perp \Lambda_\mu^k(\mathbf{R}) d\mathbf{R}, \quad (52)$$

$$\rho_\beta = \sum_p e_p \sum_{n=1}^N w_{p,n}(t) \Lambda_\beta^k(\mathbf{R}_n), \quad (53)$$

and we receive a system of equation of the form

$$P\hat{\phi} = \rho. \quad (54)$$

The so-called Poisson matrix $P$ does not depend on time and thus only has to be calculated once at the beginning of the simulation and can be used afterward for each time step, whereas the right-hand-side $\rho$ can be interpreted as the charge distribution of the particles and needs to be recalculated at each step according to the particle positions. Once having calculated $\rho$, $\hat{\phi}$ can be easily achieved from Eq. (54).

In the case of our modified polarization equation (29), the derivation is very similar. The key difference is that the spline derivatives of the matrix (52) are replaced by the splines itself—also called the mass matrix—and the whole term is multiplied by $\bar{k}_\perp^2$. The potential is calculated on defined field grid points, and then, the $E$-field is projected back to each particle position via the derivatives of the B-spline functions.

Since our discrete PIC algorithms are derived from the discrete Lagrangian, the PIC scheme fully conserves energy in the limit of $dt = 0$.[26]

## B. Sheath model

The scales present within the physical Debye sheath that builds up at the plasma-wall boundary are quite different from the ones that can be treated in gyrokinetic models. Also, increasing spatial and temporal resolution solely for the sheath is computationally not viable. But to be able to simulate open field line plasmas, it is necessary to take heat and particle fluxes toward the wall into account. To resolve this dilemma, an appropriate sheath model is required to model the effects of a Debye sheath, without actually resolving it. A suitable model therefore is the so-called logical sheath boundary conditions.

The setup of the logical sheath used in this work is generally based on the model described in Parker et al.[21] This model was originally developed for fully kinetic 1D2V PIC simulations. Recent works[18,20] also implemented Parker's logical sheath boundary conditions for parallel heat flux studies in gyrokinetic 1D1V continuum-code simulations. We will also implement these boundary conditions for our 1D1V PIC simulations.

In this model, the wall is regarded as insulating and therefore the total parallel current to the wall is set to zero ($j_\parallel = 0$) at every moment in time and on each field line. Due to this, these boundary conditions are also called insulating boundary conditions.

Physically, incident ions that flow toward the sheath are accelerated by the dropping sheath potential. Electrons, however, can only be absorbed if their velocity is high enough to overcome the sheath potential drop at the wall. Slower electrons will be reflected back into the domain. By setting $j_\parallel = 0$ within the logical sheath model, only the fastest electrons are allowed to overcome the sheath potential drop to balance the number of ions hitting the wall. The ions instead can freely exit the system. The sheath potential can be calculated by the velocity of the slowest electron, the electron cut-off velocity $v_{ce}$, still exiting the domain according to the formula

$$\delta\phi = \phi_{sh} - \phi_w = \frac{m}{2e} v_{ce}^2, \quad (55)$$

with $\phi_{sh}$ being the sheath potential and $\phi_w$ the wall potential (which is set to $\phi_w = 0$ for a grounded wall). By using z as the spatial coordinate, the $j_\parallel = 0$ condition at the sheath position $z_{sh}$ can be written as

$$\Sigma_i q_i \int_0^\infty f_i(z_{sh}, v_\parallel, t) v_\parallel dv_\parallel = e \int_{v_{ce}}^\infty f_e(z_{sh}, v_\parallel, t) v_\parallel dv_\parallel. \quad (56)$$

Due to the total absorption of ions and the partial absorption of electrons at $z_{sh}$, the distribution functions for electrons and ions at the wall become

$$f_i(z_{sh}, -v_\parallel, t) = 0 \quad v_\parallel > 0, \quad (57)$$

$$f_e(z_{sh}, -v_\parallel, t) = \begin{cases} 0 & v_\parallel > v_{ce}, \\ f_e(z_{sh}, v_\parallel, t) & v_{ce} > v_\parallel > 0, \end{cases} \quad (58)$$

with $-v_\parallel$ describing velocities of particles moving away from the wall, back into the system. Looking at the boundary condition on the ions in Eq. (57), it is important to mention that the formulation would not be sufficient to treat the cold ion case. Here, no force would act on the ions to accelerate them toward the wall. From Eq. (57), the ions will also not feel the force from the potential drop across the logical sheath and thus the Bohm criterion will not be satisfied. Future work could consider implementing these effects in the equations.

In the presented 1D1V sheath model, the $j_\parallel = 0$ condition is valid on each field line. In the 3D case, however, a current flow into the wall at one point and an outflow at a different point are allowed.

The total integrated flow in and out of the wall, however, still needs to cancel, and for the total parallel current into the wall, $j_\parallel = 0$ must hold.

### C. Algorithm

In a PIC code, the logical sheath model can be numerically implemented according to the following algorithm (based on Parker et al.[21]):

(1) Advance particle trajectories
(2) At each time step, count the number of electrons $n_e$ and ions $n_i$ that hit the wall
(3) Compare $n_e$ and $n_i$
  (a) If $n_i \leq n_e$ (probable condition)
    - Order all $n_e$ electrons according to velocity
    - Let the fastest $n_i$ electrons and $n_i$ ions leave the domain
    - Reflect the remaining $n_e - n_i$ slow electrons
  (b) If $n_e \leq n_i$ (very improbable condition)
    - Order all $n_i$ ions according to velocity
    - Let the fastest $n_e$ ions and $n_e$ electrons leave the domain
    - Reflect the slowest $n_i - n_e$ ions

Figure 1 shows a graphical representation of how the algorithm works. The sheath potential can then be calculated from the electron cut-off velocity $v_{ce}$ according to Eq. (55). The $n_e \leq n_i$ case is very improbable and does in principle not appear; thus, the (b) part is rather for completeness. Nevertheless, it is important to remark that writing the algorithm in the current form meets the $j_\parallel = 0$ condition.

## IV. SIMULATION SETUP
### A. Initial conditions

Electrons and ions are initially set to a fixed spatial and velocity distribution to achieve comparability with previous studies. However, the simulation results are mostly insensitive to the initial background. Pitts et al.,[14] for example, ran their fully kinetic PIC simulations initially with a weaker pre-ELM source to achieve a quasisteady state.

#### 1. Electron initial conditions

The initial electron distribution function is determined by

$$f_{e0}(z, v_\parallel, T_{e0}) = n_{e0}(z) F_M(v_\parallel, T_{e0}), \quad (59)$$

with $F_M(v, T_{p0}) = \frac{1}{\sqrt{2\pi T_{p0}/m_p}} \exp\left(\frac{-m_p v_\parallel^2}{2T_{p0}}\right)$, the Maxwellian distribution in 1D for species $p$ (in this case, $p$ stands for electrons). The initial temperature is set to $T_{e0} = 75$ eV, and the electron density profile (in $10^{13}$ cm$^{-3}$) has the form

$$n_{e0}(z) = 0.7 + 0.3\left(1 - \left|\frac{z}{L}\right|\right) + 0.5\cos\left(\frac{\pi z}{L_s}\right) H\left(\frac{L_s}{2} - |z|\right), \quad (60)$$

with $L$ being half the size of the simulation domain $[-L; L]$, $L_s$ the size of the source, and $H(.)$ the Heaviside step function.

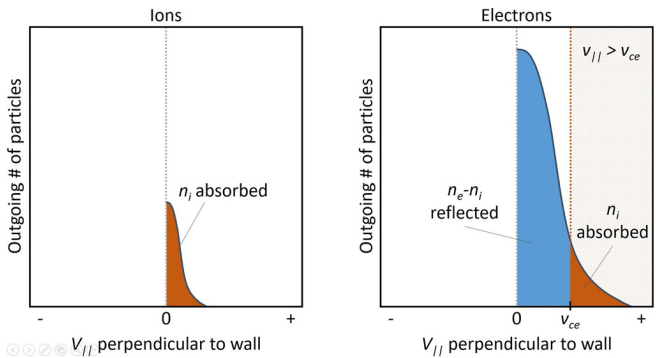

**FIG. 1.** Schematic representation of the number of outgoing ions and electrons at the sheath position as a function of $v_\parallel$ (with different scales). As described in the algorithm in Sec. III C, all $n_i$ ions are absorbed by the wall, whereas only the fastest $n_i$ electrons are absorbed. The $n_e - n_i$ slower electrons are reflected back into the plasma.

#### 2. Ion initial conditions

In the ion case, the initial distribution function is modeled as a combination of left and right half-Maxwellian distribution functions

$$\begin{aligned} F_L(z, v_\parallel, T_{i0}) &= 2n_{i0}(z) F_M(v_\parallel, T_{i0}) H(-v_\parallel), \\ F_R(z, v_\parallel, T_{i0}) &= 2n_{i0}(z) F_M(v_\parallel, T_{i0}) H(v_\parallel), \end{aligned} \quad (61)$$

where the initial ion density $n_{i0}$ is set equal to the initial electron density $n_{e0}$. Depending on the position of the particle, the ion distribution function is defined as

$$f_{i0}(z, v_\parallel, T_{i0}) = \begin{cases} F_L & z < -\frac{L_s}{2}, \\ \left(\frac{1}{2} - \frac{z}{L_s}\right) F_L + \left(\frac{1}{2} + \frac{z}{L_s}\right) F_R & -\frac{L_s}{2} < z < \frac{L_s}{2}, \\ F_R & \frac{L_s}{2} < z. \end{cases} \quad (62)$$

The ion temperature profile is determined as (in electronvolts)

$$T_{i0}(z) = 100 + 45\left(1 - \left|\frac{z}{L}\right|\right) + 30\cos\left(\frac{\pi z}{L_s}\right) H\left(\frac{L_s}{2} - |z|\right). \quad (63)$$

The initial conditions for electrons and ions are equal to the setup in Pan et al.,[20] which, in general, is equal to the setup chosen in Shi et al.,[18] except for the ion particle distribution. Here, initially, the ion density profile is defined in a way that electrons are distributed according to the Boltzmann relation to minimize the excitation of high-frequency shear Alfvén waves in the electrostatic limit. But similar to Pan et al.,[20] also in our simulations, by setting the initial ion density equal to the initial electron density, no numerical problems occur. As stated in Pan et al.,[20] the $\cos(\pi z/L_s)$ combined with the Heaviside function $H(L_s/2 - |z|)$ used in the initial and source term profiles can be replaced by an exponential function $\sqrt{2/\pi} \exp(-(\pi z/L_s)^2/2)$, which changes the simulation results only marginally.

### B. ELM and inter-ELM phase

The setup and parameters of the ELM and subsequent inter-ELM phase are based on a simplified experimental case from the JET Tokamak.[16] The model, in general, represents a plasma blob exhausted in an ELM crash and is modeled via a hot electron-deuterium source at the SOL midplane. In a 1D setup, this source is positioned at the center of the domain $[-L,L]$ with two divertor plates at each end as boundaries. After the 200 $\mu$s long ELM phase, an inter-ELM phase follows with a colder and weaker source. The source function $S_{ELM}$ depends on position $z$, time $t$, and velocity $v_\parallel$ and can be written as

$$S_{ELM}(z, v_\parallel, t) = g(t) S(z) F_M(v_\parallel, T_s(t)), \quad (64)$$

with

$$S(z) = S_0 \cos\left(\frac{\pi z}{L_s}\right) H\left(\frac{L_s}{2} - |z|\right), \quad (65)$$

$$g(t) = \begin{cases} 1 & 0 \leq t \leq 200 \mu s, \\ 1/9 & 200 \mu s < t, \end{cases} \quad (66)$$

$$T_e = \begin{cases} 1500 \text{ eV} & 0 \leq t \leq 200 \mu s, \\ 210 \text{ eV} & 200 \mu s < t, \end{cases} \quad (67)$$

$$T_i = \begin{cases} 1500 \text{ eV} & 0 \leq t \leq 200 \mu s, \\ 260 \text{ eV} & 200 \mu s < t. \end{cases} \quad (68)$$

The particle source intensity is set to $S_0 = A n_{ped} c_{s,ped}/L_s = 9.066 \times 10^{17} \text{cm}^{-3} \text{s}^{-1}$ and scales with the pedestal density and temperature, where the proportionality constant $A$ is set to $1.2\sqrt{2} \approx 1.7$. Within the code, the particle source is implemented via a Monte Carlo generation of particles, according to the described source function.

#### 1. Relevant parameters

The most relevant simulation parameters are displayed in Table I.[16] We performed simulations with varying numbers of field grid cells, particles per cell, and perpendicular wave numbers to test convergence. To achieve a high enough resolution for the relevant sheath parameters (sheath potential, heat flux, etc.), we chose a minimum of $\sim$10 000 particles per cell. The physical results are still achievable for a lower number of particles per cell; however, the signal to noise ratio is affected. Since we did not want to discuss numerical convergence in detail, we chose a rather high number of markers to ensure good resolution. For a high enough spatial resolution of the fields in the z-direction, $n_z \geq 16$ grid cells should be set. In the following, we will show results for a run with $\sim$100 000 particles per cell, $n_z = 32$ and $k_\perp \rho = 0.2$. We will also show what effect varying $k_\perp \rho$ (0.05–1.0) has on the simulation results. The high number of particles per cell of this simulation is excessive, but since it can be run within hours, we did not see the necessity go to lower resolution. In general, one has to keep in mind that for full-f codes, a higher number of particles per cell (>1000) is required than for delta-f. Once going to 3D2V, we plan to reduce the number of particles per cell and apply noise-reduction techniques to save computational time, if necessary.

**TABLE I.** Simulation parameters used for 1D1V ELM heat pulse simulations.

| Parameter | Value | Description |
| --- | --- | --- |
| $2L$ | 80 m | Length of the simulation domain |
| $L_s$ | 25 m | Length of the source region |
| $t_{ELM}$ | 200 $\mu$s | Duration of the ELM phase |
| $\tau_i$ | 149 $\mu$s | Ion transit time ($L/c_{s,ped}$) |
| $\tau_e$ | 2.5 $\mu$s | Electron transit time ($L/v_{te,ped}$) |
| $T_{ped}$ | 1500 eV | Ion and electron temperature at the ELM pulse |
| $S_0$ | $9.066 \times 10^{17} \text{cm}^{-3}\text{s}^{-1}$ | Particle source intensity |
| $k_\perp \rho$ | 0.2 | Perpendicular wave number |
| $B$ | 2 T | B-field strength in the parallel direction |

### V. SIMULATION RESULTS

To extensively study the plasma behavior during the ELM phase and beyond, the spatial profiles at the end of the ELM phase and the time-dependent heat flux toward the divertor plates are investigated. Additionally, the sheath potential $\phi_{sh}$ development over time will be studied. Since the problem is symmetric, $\phi_{sh}$ can be measured arbitrarily at the right or left boundary.

### A. ELM phase spatial profiles

The spatial profiles of interest for the heat flux problem for species $p$ are density $n_p$, parallel particle flux $\Gamma_p$, parallel temperature $T_{\parallel,p}$, and parallel heat flux $Q_p$ and are defined as

$$n_p = \int_{-\infty}^{\infty} f_p dv_\parallel, \quad (69)$$

$$\Gamma_p = \int_{-\infty}^{\infty} f_p v_\parallel dv_\parallel, \quad (70)$$

$$Q_p = \frac{1}{2} m_p \int_{-\infty}^{\infty} f_p v_\parallel^3 dv_\parallel + T_\perp \int_{-\infty}^{\infty} f_p v_\parallel dv_\parallel, \quad (71)$$

$$T_{\parallel,p} = \frac{1}{n_p} m_p \int_{-\infty}^{\infty} f_p (v_\parallel - \langle v_\parallel \rangle_p)^2 dv_\parallel, \quad (72)$$

with $f_p$ being the particle distribution function and the perpendicular temperature set to the pedestal temperature $T_\perp = T_{ped} = 1500$ eV (equal to that reported in previous studies).

For all figures within this section, we calculated the values for the $n_z$ grid cells and performed a cubic spline interpolation on 128 diagnostics cells to achieve a smooth shape of the profiles. In Fig. 2, these profiles are plotted for $k_\perp \rho = 0.2$ right before the source is switched off and the system comes into the post-ELM phase (at 200$\mu$ s).

It is important to mention that in the upper left plot in Fig. 2, the gyrocenter density is plotted, which means that in the case of the electrons, the polarization term of Eq. (29) is added. The overlap of the ion and electron gyrocenter density is a good indicator for the accuracy of the simulation and shows that the electrons are mostly bound to the ions and are transported toward the boundary with the same flux rate. This is in very good agreement with the results seen in Pan et al.[20] The slight excess of electrons directly at the boundaries is due to the logical sheath boundaries that lead to an absorption (reflection) of ions (electrons) by the wall.

Also, the other spatial profiles are in good agreement with the results of Pan et al.[20] In the case of the temperature, the electron temperature is slightly higher than in the Pan result, but still the overall lower value for electrons compared to ions has the same origin of selective loss of high-energy electrons to the wall. This slightly higher value is an indicator that in our PIC algorithm, we allow higher energy electrons in our particle sources. Using a lower limit for the electron velocity in the source distribution function also decreases the electron temperature. In the case of the parallel particle and heat flux, the electron fluxes oscillate around the ion profiles, due to PIC-inherent noise effects, but still a good overlap could be achieved. Increasing the number of particles further or averaging over the diagnostics cells can further flatten the profiles and leads to a good overlap between both species.

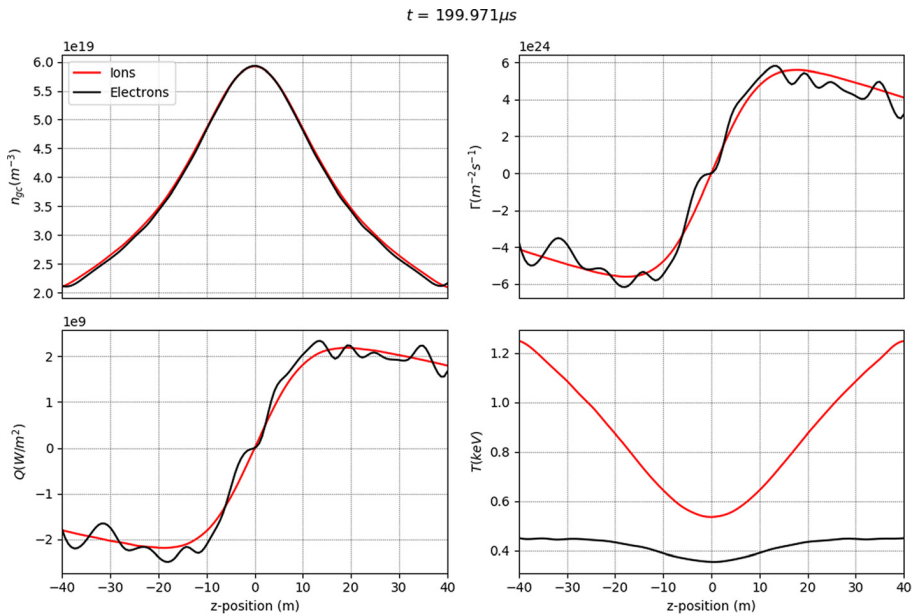

**FIG. 2.** Spatial profiles of electrons (black) and ions (red) of gyrocenter density $n_{gc}$, parallel particle flux $\Gamma$, parallel heat flux $Q$, and parallel temperature $T_\parallel$ within 1D simulation domain for $k_\perp \rho = 0.2$. The snapshot of the profiles is taken shortly before the end of the ELM phase at $200 \mu s$.

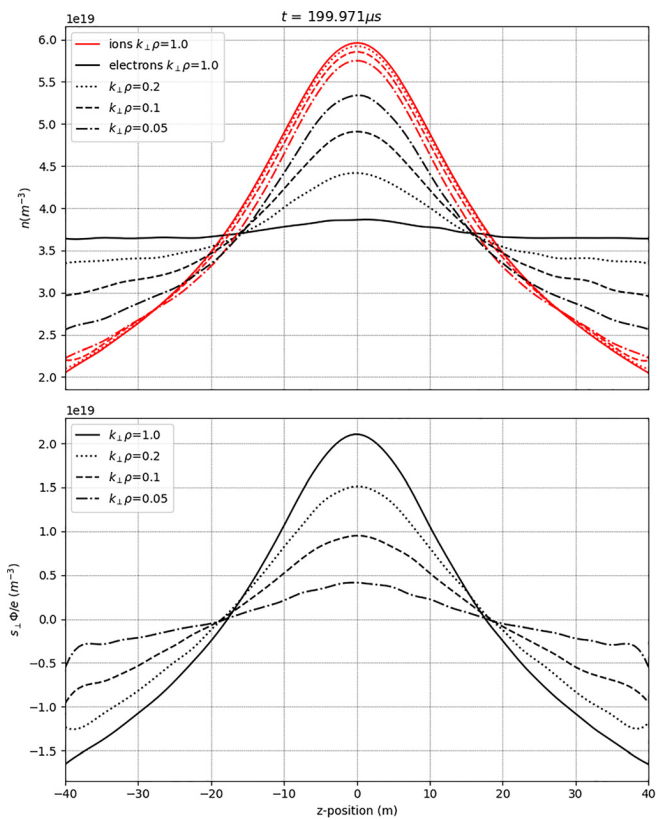

**FIG. 3.** Spatial density profiles for electrons (without the polarization term) and ions and the polarization $s_\perp(z)\phi(z)$ divided by $e$ for varying $k_\perp \rho = 0.05 - 1.0$.

In both cases, the drop of the fluxes toward the domain boundary is a clear indicator that after $200 \mu$ s, the system has not yet reached an equilibrium between the introduced particles from the source and the lost particles at the walls. In an equilibrated state, the flux profiles outside the source region are flat toward the domain boundary.

To show the effects of varying $k_\perp \rho$, the densities for ions and electrons (without the polarization term) together with the polarization $s_\perp(z)\phi(z)$, divided by $e$ to achieve comparability, are plotted in Fig. 3 for values of $k_\perp \rho = 0.05 - 1.0$. With decreasing $k_\perp \rho$, the ion (electron) density is decreasing (increasing). Since the electrons are the lighter and more mobile species, their density profile is affected more extensively. This in turn has the effect that the difference between the ion and electron densities becomes smaller, which can be directly seen in the decreasing polarization, where $s_\perp(z)\phi(z)/e$ is equal to the density difference. In other words, the electrons are strongly bound to the ions. This could also be seen by increasing the electric field that evolves to maintain quasineutrality of the system, which we want to neglect here. The other profiles, temperature and particle and heat flux, are hardly affected by changing $k_\perp \rho$.

### B. Divertor heat flux and sheath potential

The calculation of the parallel heat flux on the divertor targets in gyrokinetic simulations is similar to the calculation of the heat flux

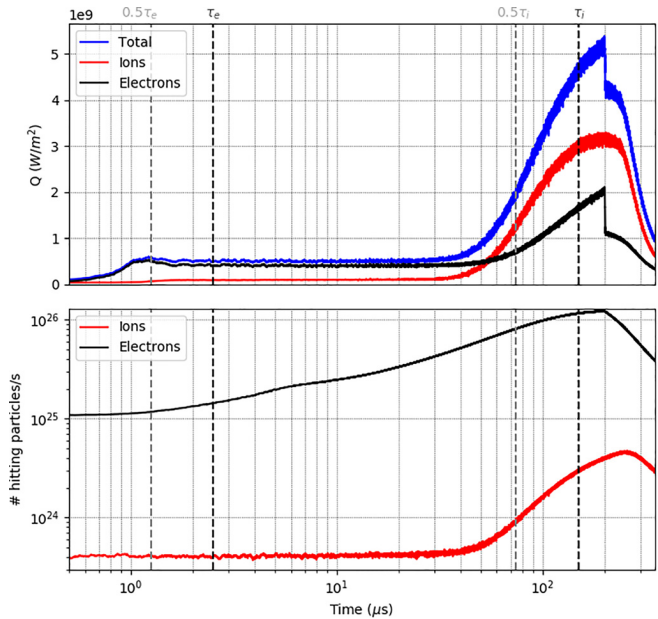

**FIG. 4.** Evolution of ion (red), electron (black), and total (blue) heat flux, according to Eq. (73), and number of hitting particles on the right sheath boundary. The (half of the) thermal ion and electron transit times $\tau_e$ and $\tau_i$ are indicated by (gray) black vertical lines.

spatial profile [see Eq. (71)], but additional sheath effects have to be taken into account to obtain

$$Q_p = \frac{1}{2} m_p \int_{v_{c,p}}^{\infty} f_p v_\parallel^3 dv_\parallel + (T_\perp + q_p \phi_{sh}) \int_{v_{c,p}}^{\infty} f_p v_\parallel dv_\parallel. \quad (73)$$

Since the electrons that are reflected back into the plasma are not taken into account, the lower boundary of the integrals is $v_{c,p} = \sqrt{\max(-2q_p \phi_{sh}/m_p, 0)}$, which is based on the definition of the cut-off velocity defined in Eq. (55). In the less probable case that more ions than electrons hit the divertor, the same would hold for the reflected ions. The additional $\phi_{sh}$ term is added to account for the acceleration (deceleration) of outgoing ions (electrons) by the sheath.

For all graphs within this section, a moving average of about 50 time steps ($\sim 0.1 \mu s$) was chosen to decrease the number of required particles per cell to achieve fast simulations and results with low enough noise. In Fig. 4, the heat flux on the right sheath boundary, as well as the number of particles that hit this boundary, is shown for the same $k_\perp \rho = 0.2$ run as in Sec. V A. Since logical sheath boundary conditions are applied, most of the electrons are reflected and only as many electrons as ions eventually leave the domain. In the shown simulation, at each point in time, more electrons than ions are hitting the wall, and thus, the number of electrons and ions that are absorbed by the wall has to be equal to the number of ions that hit the wall.

The parallel electron heat flux rises quickly for values smaller than $0.5\tau_e = 0.5L/\sqrt{T_{ped}/m_e}$, which results from the first fast electrons of the hot ELM source hitting the wall. In the following, the electron heat flux slightly decreases before remaining constant at $\sim 0.4 \times 10^9 \mathrm{W/m^2}$ until $\sim 0.5\tau_i$. At the same time, the ion heat flux initially rises very slightly and remains at a rather low value of $\sim 0.1 \times 10^9 \mathrm{W/m^2}$, which leads to a constant total heat flux of $\sim 0.5 \times 10^9 \mathrm{W/m^2}$ until shortly before $0.5\tau_i$. The slight increase (decrease) of the ion (electron) heat flux at $\sim 0.5\tau_e$ is due to the sheath potential that builds up at that time (see Fig. 5). Previous studies with continuum codes (Fig. 2 in Pan et al.[20] and Fig. 3 in Shi[19]) show a similar increase in the total heat flux to $\sim 0.5 \times 10^9 \mathrm{W/m^2}$ in this time period; however, the ion heat flux increases as much as the electron flux drops at $0.5\tau_e$ and remains higher for the rest of their simulations. In the case of simulations with a fully kinetic PIC code (Fig. 2 in Havlíčková et al.[16]), the electron flux remains above the ion flux for this period, similar to our case.

At about half of the ion transit time, when the first fast ions from the ELM arrive, both electron and ion heat fluxes increase steeply until the ELM crash at $200 \mu s$, where a peak total heat flux of $\sim 5.1 \times 10^9 \mathrm{W/m^2}$ is reached. At this point, the electron flux drops immediately, whereas the ion response is significantly slower and leads to a deferred drop. Comparing this period again to the previous simulations with continuum codes shows good qualitative agreement. The peak total heat flux for electrons and ions and thus the total heat flux, however, are slightly higher ($\sim 5.1 \times 10^9 \mathrm{W/m^2}$ compared to $\sim 4.1 \times 10^9 \mathrm{W/m^2}$).

This difference is mainly caused by the differing source implementation of our PIC vs the continuum codes of previous studies. By cutting-off the electron velocities at lower values or narrowing the velocity distributions of the particle sources, the same maximum values of the heat fluxes can be constructed. In the case of the fully kinetic PIC code, the peak total flux is very similar to our result, but the peak ion (electron) flux is significantly higher (lower) than in our case, which can be due to the different physics solved.

Integrating the fluxes over time delivers the total ELM energy transported to the divertor for each species. In the case of the ions (electrons), this is 66.6% (33.4%) of the total ELM energy and again is very similar to the values achieved with the previous continuum code simulations.

The heat fluxes on the domain wall include deceleration (acceleration) of electrons (ions), which the SOL heat fluxes in Sec. V A do not include. The reason for the strong increase in the heat flux at $\sim 0.5\tau_i$ can be understood by looking at the number of hitting particles in Fig. 4. This clearly shows the steep increase in incoming ions at the sheath boundary. The still rising number of incident ions after $200 \mu s$ is also a good indicator for the deferred ion heat flux drop. However, in the case of the electrons, the increase in hitting particles as expected starts already at $\sim 0.5\tau_e$ and immediately decreases after switching off the ELM source.

To get an even more complete physical picture, in Fig. 5, the time-dependent evolution of the sheath potential $\phi_{sh}$ is shown. At the beginning, $\phi_{sh}$ is only determined by the cold initial distribution. Around $\sim 0.5\tau_e$, the potential rapidly rises to $\sim 3$ keV, due to the arriving suprathermal electrons from the ELM source. Until the arrival of the suprathermal ions at $\sim 0.5\tau_i$, the potential stays mainly constant. The large sheath potential leads to a deceleration and reflection of electrons at the divertor. The majority of electrons are prevented from hitting the divertor, and thus, the increase in the electron flux is stopped at $\sim 0.5\tau_e$ and slightly inverted. After the arrival of the suprathermal ions, the sheath potential drops steadily and allows an increase in both the ion and electron flux. Compared to Pan et al.,[20] the qualitative behavior is very similar, with the difference that the increase in the previous study is even steeper and the absolute value is slightly lower ($<3\,\text{keV}$). This again is an indicator that the ELM source in our PIC approach introduces faster electrons into the system.

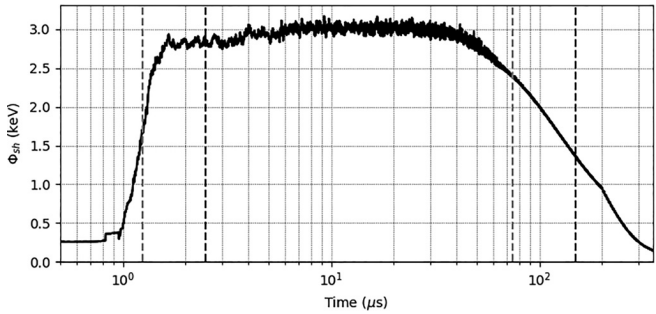

**FIG. 5.** Time-dependent evolution of the potential at the right sheath boundary. The (half of the) thermal ion and electron transit times $\tau_e$ and $\tau_i$ are indicated by (gray) black vertical lines.

As mentioned before, the quantitative differences of our results compared to the previous continuum code results from Pan et al.[20] and Shi[19] are mainly based on our PIC-specific implementation of the velocity distribution functions. We assume that the sources we implemented allow faster particles than the sources used in the mentioned studies. In Fig. 6, we therefore plotted the heat flux on the divertor for a simulation with a lower maximal limit for the velocity distribution of the particle sources. Compared to our previous simulation runs, we thus initialize electrons and ions with a lower maximum velocity of $v_\parallel \leq 2.2 v_{\text{th}} = 2.2\sqrt{k_B T_p/m_p}$ (vs $v_\parallel \leq 3.7 v_{\text{th}}$), or with $f_p(v_\parallel)/f_{p,\max} = 10\%$ (vs 0.1%). The qualitative behavior of the heat flux compared to the previous run is still maintained; however, the peak values are decreased, since the velocity of the fastest particles is limited.

A very similar picture can be achieved by narrowing the distribution function and thus also limiting the very high velocity particles. In a similar way, the sheath potential $\phi_{\text{sh}}$ reaches lower peak values by setting a lower velocity limit. Compared to the sheath parameters, the spatial profiles, however, are only slightly affected, since the majority of particles are below the velocity limit and the effects of the fastest particles are less crucial.

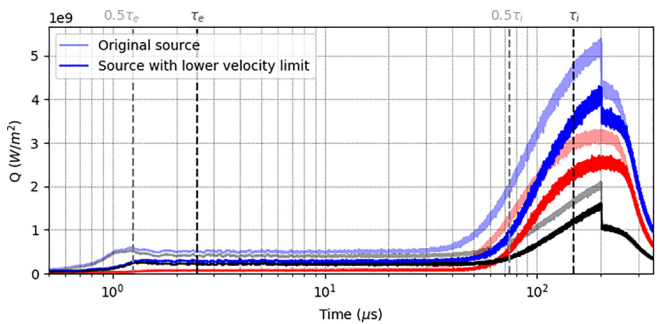

**FIG. 6.** Comparison of the heat flux on the divertor for the originally implemented source (transparent, as already shown in the upper plot of Fig. 4) and for a source with a decreased limit for the maximum of the velocity distribution (opaque), where $v_\parallel \leq 2.2 v_{\text{th}}$ is set. Similar to Fig. 4, the evolution of ion (red), electron (black), and total (blue) heat flux is shown for both source types.

## VI. CONCLUSIONS

We introduced the equations and numerical techniques used in PICLS, a gyrokinetic PIC code specifically designed for open-field line simulations in the SOL. To take into account the large amplitude fluctuations, which are present in edge and SOL plasma, an electrostatic full-f model with a linearized polarization equation was studied. Additionally, logical sheath boundary conditions were implemented to simulate particle loss to and reflection from the domain walls and the emerging sheath potential without having to resolve down to the electron Debye length.

To test the implementation, we implemented the well-studied 1D ELM heat pulse problem with a central heat pulse that propagates along the field line toward the divertor target. The results for the heat flux on the divertor targets are consistent with previous fully kinetic continuum, PIC, fluid, and gyrokinetic continuum simulations. Our results are in-line with previous findings: the ELM heat flux loading on the divertor occurs mainly on the ion transit time scale; a strong negative potential builds on the electron transit time and confines the majority of electrons; the asymmetric heat flux on the divertor of electrons and ions is caused by deceleration (acceleration) of electrons (ions) by the sheath. The differences in the peak values compared to previous gyrokinetic continuum simulations are due to the code specific particle source implementation. Additionally to previous findings, we showed that varying $k_\perp \rho$ mainly affects the potential and therewith the ion polarization. The heat flux on the divertor and the sheath potential, however, are hardly affected (at least for values of $1.0 \geq k_\perp \rho \geq 0.05$).

By implementing the nonlinear Poisson equation (29) in our model, we assumed only a single $k_\perp$ mode. A more accurate calculation of the potential and accounting for the coupling of several $k_\perp$ modes could be achieved by using Eq. (25) in the future instead. Coupling of different $k_\perp$ modes is decisive, when going toward more realistic edge plasma systems or real experimental machines like LAPD. For this, we already started the extension of PICLS toward a 3D gyrokinetic model.

As suggested in previous fully kinetic PIC simulations,[14,31] for the SOL transport, collisions can also be important. Thus, the implementation of a collision operator into PICLS is required for the extension toward higher dimensions.

### ACKNOWLEDGMENTS

The authors would like to thank Stephan Brunner and Laurent Villard from the Swiss Plasma Center (SPC, Lausanne) for their collaboration and support. Numerical simulations were performed on the Marconi supercomputer within the framework of the PICLS project. This work was carried out within the framework of the EUROfusion Consortium and received funding from the Euratom research and training program 2014–2018 and 2019–2020 under Grant Agreement No. 633053, for the No. EF WP32-ENR-MPG-04 (2019/2020) project "MAGYK." The views and opinions expressed herein do not necessarily reflect those of the European Commission.